\documentstyle[mncite,psfig]{mn}
\title[Alignment of Clusters]{The Alignment of Clusters using Large Scale Simulations}
\author[L.I. Onuora \& P.A. Thomas]{Lesley I. Onuora \thanks{email: lonuora@astr.cpes.susx.ac.uk}\& Peter A. Thomas \\
Astronomy Centre, University of Sussex, Falmer, Brighton BN1 9QJ}
\date{accepted --------------. Received ------------; in original form-------}

\begin{document}
\maketitle

\begin{abstract}
The alignment of clusters of galaxies with their nearest neighbours and between clusters within a supercluster is investigated using simulations of $512^{3}$ dark matter particles for $\Lambda$CDM and $\tau$CDM cosmological models. Strongly significant alignments are found for separations of up to $15 h^{-1}$Mpc in both cosmologies, but for the $\Lambda$CDM model the alignments extend up to separations of $30h^{-1}$Mpc. The effect is strongest for nearest neighbours, but is not significant enough to be useful as an observational discriminant between cosmologies. 
As a check of whether this difference in alignments is present in other cosmologies, smaller simulations with $256^{3}$ particles are investigated for 4 different cosmological models. Because of poor number statistics, only the standard CDM model shows indications of having  different alignments from the other models.

\end{abstract}


\section{INTRODUCTION}

Differing claims have been made as to the scale and significance of alignments of clusters of galaxies. Binggeli (1982) was the first to point out this effect. He found that not only were clusters aligned with their nearest neighbours out to a separation of about $15h^{-1}$Mpc (where 
$h=H_{0}/100$ $\mathrm{km}$ $\mathrm{s}^{-1} \mathrm{Mpc}^{-1}$)  but that the orientation of a cluster was also related to the distribution of all surrounding clusters up to  a separation of about $50h^{-1}$Mpc. Since then the effect has been studied by many authors (Struble \& Peebles 1985; Flin 1987; Rhee \& Katgert 1987; West 1989;  Ulmer, McMillan \& Kowalski 1989; West, Dekel \& Oemler 1989; Fong, Stevenson \& Shanks 1990; Plionis 1994; Martin et al. 1995; Splinter et al. 1997) but with conflicting results. A number of factors needed to be considered, particularly in the  observational studies, before definite conclusions could be drawn as to the reality of the effect.

One major problem is that of determining the position angles of the clusters. For this reason Flin (1987) used several independently determined position angles in his study, and Rhee and Katgert (1987) developed a more objective semi-automatic procedure to determine position angles. As a result both studies found some support for the alignment effects found by Binggeli (1982). However
Martin et al. (1995) pointed out that a fundamental problem is that of the number of galaxies being too few to trace the cluster potential adequately. In addition, many of the studies have used clusters from catalogues (e.g. Abell and Lick) which are known to contain systematic biases (e.g. Lumsden et al. 1992). Using instead  a statistical search of the Edinburgh-Milano cluster redshift survey, Martin et al. (1995) found no statistically significant evidence of cluster alignments. 

There would be considerable advantages in using X-ray data rather than optical data since hot gas traces the cluster potential directly and reduces the problem of confusion of cluster membership along the line of sight. Ulmer et al. (1989) used 45 X-ray clusters and found no significant effect. However, Rhee, van Haarlem \& Katgert (1992) found that, using combined optical and X-ray data, clusters do tend to point towards neighbouring clusters if they are members of the same supercluster. 

Large scale simulations of clusters can provide large data sets of clusters for different cosmological models without observational biases and for which the orientation of the semi-major axis can be derived in a straightforward manner. A comparison between the simulations and observations can reveal whether the observations are consistent with any of the cosmologies. Whether the observations could be used to discriminate between models depends on the statistical significance of the difference in alignments between cosmological models, which can also be obtained from the simulations. If the alignments are sufficiently different in different models, then large observational data sets of cluster position angles could be used to test for the cosmology. The best data presently available is that of Plionis (1994) who estimated position angles for 637 clusters in contrast to the very much smaller numbers of clusters of about 50 or less used in other observational studies of alignments. Splinter et al. (1997) used simulations of $128^{3}$ particles to study the ellipticity and orientation of clusters of galaxies in N-body simulations for different cosmological models (ie. different values of density parameter, $\Omega_{o}$, and initial power spectra). The box sizes varied from 110 to 300$h^{-1}$Mpc depending on the initial spectral index and the evolutionary stage. They found significant alignments for all spectra at separations less than $15h^{-1}$Mpc and that differences in $\Omega$ had no measurable effect. In the present work cluster alignments in different cosmologies will be looked for using significantly larger simulations having $512^{3}$ and $256^{3}$ particles.

Explanations of the alignment of galaxy clusters, if this effect exists, fall into 2 main categories. Tidal distortion due to interactions between clusters of galaxies has been proposed as the origin of the ellipticity of clusters (Binney \& Silk 1979) and also of alignment (Salvador-Sol\'{e} \& Solanes 1993),
although Dekel, West \& Aarseth (1984) had earlier found that tidal interactions
do not produce alignments and favoured instead an intrinsic origin for cluster anisotropies. Using N-body simulations they found that alignments only occurred in a 'top down' scenario in which superclusters collapsed from excess fluctuations on large scales rather than hierarchical clustering from fluctuations on smaller scales. However, other simulations of hierarchical clustering (e.g. White 1976; Cavaliere et al. 1986) have modelled the clumpy, non-spherical nature of clusters formed via subclustering. In line with this, van Haarlem \& van de Weygaert (1993) found for cold dark matter (CDM) models that clusters are elongated in the direction from which the last subcluster fell into the cluster. West, Jones \& Forman (1995) proposed that cluster formation proceeds by the merging of subclusters along large-scale filamentary features in the matter distribution. Thus the initial concern that the more accepted CDM models may be ruled out by the observed alignments in favour of a top down pancake scenario may be unfounded and cluster alignments may in fact be expected irrespective of the cluster formation model (Plionis 1994)

In the present work large scale simulations of two CDM models are analysed for cluster alignment effects both among nearest neighbours and between a cluster and its neighbours within a supercluster. This is extended to a total of 4 different models using smaller simulations. The aims are to confirm the scale and significance of any alignments and to look for any significant differences between different cosmological models. The details of the simulations and cluster catalogues are given in Section 2, the method and results of searching for alignments are given in Section 3, and the results are discussed in Section 4.

\section{SIMULATIONS AND CLUSTER IDENTIFICATION}

The box sizes used are  $479.0h^{-1}$Mpc and $320.6h^{-1}$Mpc and contain $512^{3}$ dark matter particles. Two different cosmological models with CDM power spectrum are investigated: a flat model with $\Omega$=0.3 and an $\Omega$=1 model ($\tau$CDM) set to have the same power spectrum as the $\Omega$=0.3 model. Both models have the same spectral shape parameter $\Gamma$ of 0.21 and the amplitude of primordial fluctuations is normalized so that the models reproduce the observed abundance of rich clusters at the present time as in Jenkins et al. (1998)

As explained in Thomas et al. (1998), the cluster catalogues were obtained from the simulation data by defining clusters in terms of an overdensity relative to the critical density. The minimal spanning tree was found for all particles with overdensities greater than 180 and was then truncated to divide the data into clusters.The semi-major axes, $a_{1} \ge a_{2} \ge a_{3}$ of each cluster were defined in terms of the best-fitting ellipsoid, normalised such that for a uniform sphere the semi-major axes are equal to the radius. 

For most of the subsequent analysis the minimum cluster mass used was $1.8 \times 10^{14}h^{-1} \mathrm{M}_{\odot}$. This is close to the typical lower limit for the virial mass of a rich Abell cluster of about $2 \times 10^{14}h^{-1} \mathrm{M}_{\odot}$ (eg Carlberg et al. 1996). To test for the effect of cluster richness other mass limits were also tried. 
\begin{table*}
\begin{center}
\Large
\begin{tabular}{|c|c|c|c|c|ccc|} \hline
Model & Box size/ & Particle mass & No. particles & $\sigma_{8}$ & \multicolumn{3}{c}{No. of clusters in mass range/$h^{-1}\mathrm{M}_{\odot}$}\\ 
 & $h^{-1}$Mpc & $h^{-1} \mathrm{M}_{\odot}$ & & & $>9 \times 10^{13}$ & $>1.8 \times 10^{14}$ & $>2.7 \times 10^{14}$ \\ \hline
$\Lambda$CDM & $239.5$ & $6.86\times 10^{10}$ & $256^{3}$& $0.90$ & $249$ & $88$ & $37$ \\  & $479.0$  & $6.86\times 10^{10}$ & $512^{3}$& $0.90$ & $2247$ & $703$ & $308$ \\ \hline
SCDM & $239.5$ & $2.27 \times 10^{11}$ & $256^{3}$& $0.51$ & $901$ & $190$ & $57$ \\ \hline
OCDM & $239.5$ & $6.86 \times 10^{10}$ & $256^{3}$& $0.85$ & $285$ & $93$ & $41$ \\ \hline
$\tau$CDM & $239.5$ & $2.27 \times 10^{11}$ & $256^{3}$ & $0.51$ & $541$ & $137$ & $47$ \\
 & $320.6$ & $6.86 \times 10^{10}$ & $512^{3}$ & $0.51$ & $1637$ & $435$ & $162$ \\
\hline
\end{tabular}
\end{center}
\caption{Properties of the simulations for the 4 cosmological models.}
\end{table*}

Smaller simulations with a box size of $239.5h^{-1}$Mpc and containing $256^3$ dark matter particles were also used. In this case four different cosmological models were investigated to determine whether any significant differences in alignments were present for a larger range of cosmological models, the $\Lambda$CDM and $\tau$CDM models as above and additionally an open model with $\Omega=0.3$ (OCDM) and an $\Omega$=1 standard CDM model (SCDM). For the SCDM model the parameter $\Gamma$ had the value 0.5. The box sizes, particle masses and numbers of clusters in different mass ranges are summarized in Table 1 for all of the simulations. 

\section{METHOD \& RESULTS}

\subsection{Searching for Alignments using the $512^{3}$ simulations}

The analysis was first carried out for nearest neighbour clusters only using the $512^{3}$ particle simulations. For each cluster in the simulation the nearest neighbour cluster was selected. The cosine of the acute angle, $\theta$, between the cluster's major axis and the line joining its centre to that of its nearest neighbour was then found for each cluster. A normalised cumulative plot of the number of clusters as a function of $\cos\theta$ was made and compared with an isotropic distribution. A Kolmogorov-Smirnov (KS) test was carried out on the distributions to test the significance of any deviations from isotropy. The maximum values, ${D}_{\mathrm{max}}$, of the absolute difference between the two distributions was found for each model and corrected for the number of clusters, $n$, to 
$d_{\mathrm{m}} = {D}_{\mathrm{max}}/n$ for comparison between models. The KS probability $(p)$ of obtaining this deviation by chance was calculated. 

The analysis was repeated for neighbours within some maximum separation limit (sep). For comparison with other published work separation limits of 15, 30 and 60 $h^{-1}$Mpc were tried. Figures 1 and 3 show that the alignments for nearest neighbours extend up to separations of $60h^{-1}$Mpc. Inclusion of only those clusters which are more elongated (ratio of semi-major to semi-minor axis $\geq 1.5$) was also tested. Selecting elongated clusters made no great change to the alignments, only slightly improving the values of $d_{\mathrm{m}}$ (Table 2).
Additionally the alignments for \emph{all} neighbours within the same supercluster were investigated. A percolation length of $30h^{-1}$Mpc was used to define members of a supercluster. This was intermediate between the two values tried by Plionis (1994) and resulted in 47 superclusters for the $\Lambda$CDM model and 34 superclusters for the $\tau$CDM model each with between 4 and 21 cluster members. The results in Table 2 show that again the alignments extend up to $60h^{-1}$Mpc, but generally the effect is stronger for closer separations and nearest neighbours.

To test for the effect of cluster richness on the alignments, the analysis was repeated with the sample mass limits increased by a factor of 1.5 to $2.7 \times 10^{14}h^{-1}\mathrm{M}_{\odot}$. The values of $d_{\mathrm{m}}$ increased by up to 30$\%$, but because of the smaller sample numbers the significance of the deviations from isotropy were reduced. Fuller et al. (1999) found that cluster alignments with the surrounding cluster distribution may persist irrespective of cluster richness down to poor clusters (mass in the region of $10^{13} - 10^{14}\mathrm{M}_{\odot}$). To check for this, the alignments of clusters of half the mass ($9 \times 10^{13}h^{-1}\mathrm{M}_{\odot}$) with rich neighbours were also investigated. Alignments similar to those for nearest neighbours in Table 2 were found in this case, but the significance of the alignments was much larger due to the very large sample sizes.

\begin{table*}
\begin{center}
\Large
\begin{tabular}{|l|lcr||lcr|} \hline
 & \multicolumn{3}{c}{$\Lambda$CDM} &  \multicolumn{3}{c}{$\tau$CDM} \\ \cline{2-7}
 & $d_{\mathrm{m}}$ & p & n &  $d_{\mathrm{m}}$ & p & n \\ \hline
sep$<15$(NN) & $0.32$ & $5.7\times10^{-19}$ & $204$ &  0.27 & $9.9 \times10^{-12}$ & $173$ \\
sep$<30$(NN) & $0.23$ & $8.3 \times10^{-24}$ & $512$ &  $0.17$ & $2.6\times 10^{-9}$ & $366$ \\
sep$<60$(NN) & $0.17$ & $5.4 \times 10^{-18}$ & $700$ &  $0.13$ & $1.8\times 10^{-6}$ & $435$ \\
sep$<15$(Elong) & $0.35$ & $5.3 \times 10^{-14}$ & $131$ &  $0.31$ & $2.2 \times 10^{-11}$ & $130$ \\
sep$<30$(Elong) & $0.25$ & $3.1 \times 10^{-18}$ & $328$ &  $0.19$ & $ 6.0 \times 10^{-9}$ & $284$ \\
sep$<60$(Elong) & $0.18$ & $1.8 \times 10^{-3}$ & $462$ &  $0.14$ & $6.1 \times 10^{-6}$ & $339$ \\
sep$<15$(All N) & $0.24$ & $4.8 \times 10^{-20}$ & $402$ &  $0.22$ & $6.6 \times 10^{-10}$ & $220$ \\
sep$<30$(All N) & $0.15$ & $9.0 \times 10^{-23}$ & $870$ &  $0.12$ & $2.7 \times10^{-11}$ & $878$ \\
sep$<60$(All N) & $0.12$ & $2.5 \times 10^{-21}$ & $1580$ &  $0.07$ & $1.4 \times 10^{-10}$ & $2144$\\

\hline
\end{tabular}
\end{center}
\caption{Alignments for $512^{3}$ clusters at separations of 15, 30 \& 60$h^{-1}$Mpc for nearest neighbours (NN), elongated nearest neighbours only (Elong) and all neighbours within a supercluster (All N)}
\end{table*}

To see at which separations the alignments arose, the analysis was repeated for separation ranges of 15-30 and 30-60 $h^{-1}$Mpc. The results are given in Table 3 and are shown for nearest neighbours in Figures 2 and 4.
 
\begin{table*}
\begin{center}
\Large
\begin{tabular}{|l|lcr||lcr|} \hline
Range & \multicolumn{3}{c}{$\Lambda$CDM} & \multicolumn{3}{c}{$\tau$CDM} \\ \cline{2-7}
 & $d_{\mathrm{m}}$ & p & n & $d_{\mathrm{m}}$ & p & n \\ \hline
0-15(NN) & $0.32$ & $5.7 \times 10^{-19}$ & $204$ & $0.27$ & $9.9 \times 10^{-12}$ & $173$ \\
15-30(NN) & $0.17$ & $1.6 \times10^{-8}$ & $308$ & $0.07$ & $0.25$ & $193$ \\
30-60(NN) & $0.03$ & $0.99$ & $188$ & $0.10$ & $0.53$ & $69$ \\
0-15(All N) & 0.$24$ & $4.8\times10^{-20}$ & $402$ & $0.22$ & $6.6\times10^{-10}$ & $220$
\\
15-30(All N) & $0.13$ & $1.1\times10^{-8}$ & $536$ & $0.09$ & $1.1\times10^{-4}$ & $658$
\\
30-60(All N) & $0.10$ & $5.6\times10^{-6}$ & $698$ & $0.05$ & $5.3\times10^{-6}$ & $1266$ \\
\hline
\end{tabular}
\end{center}
\caption{Alignments for $512^{3}$ particles in separation ranges 0-15, 15-30 \& 30-60 $h^{-1}$Mpc for nearest neighbours (NN) and all neighbours within a supercluster (All N).}
\end{table*}

\begin{figure}
\centerline{\psfig{figure=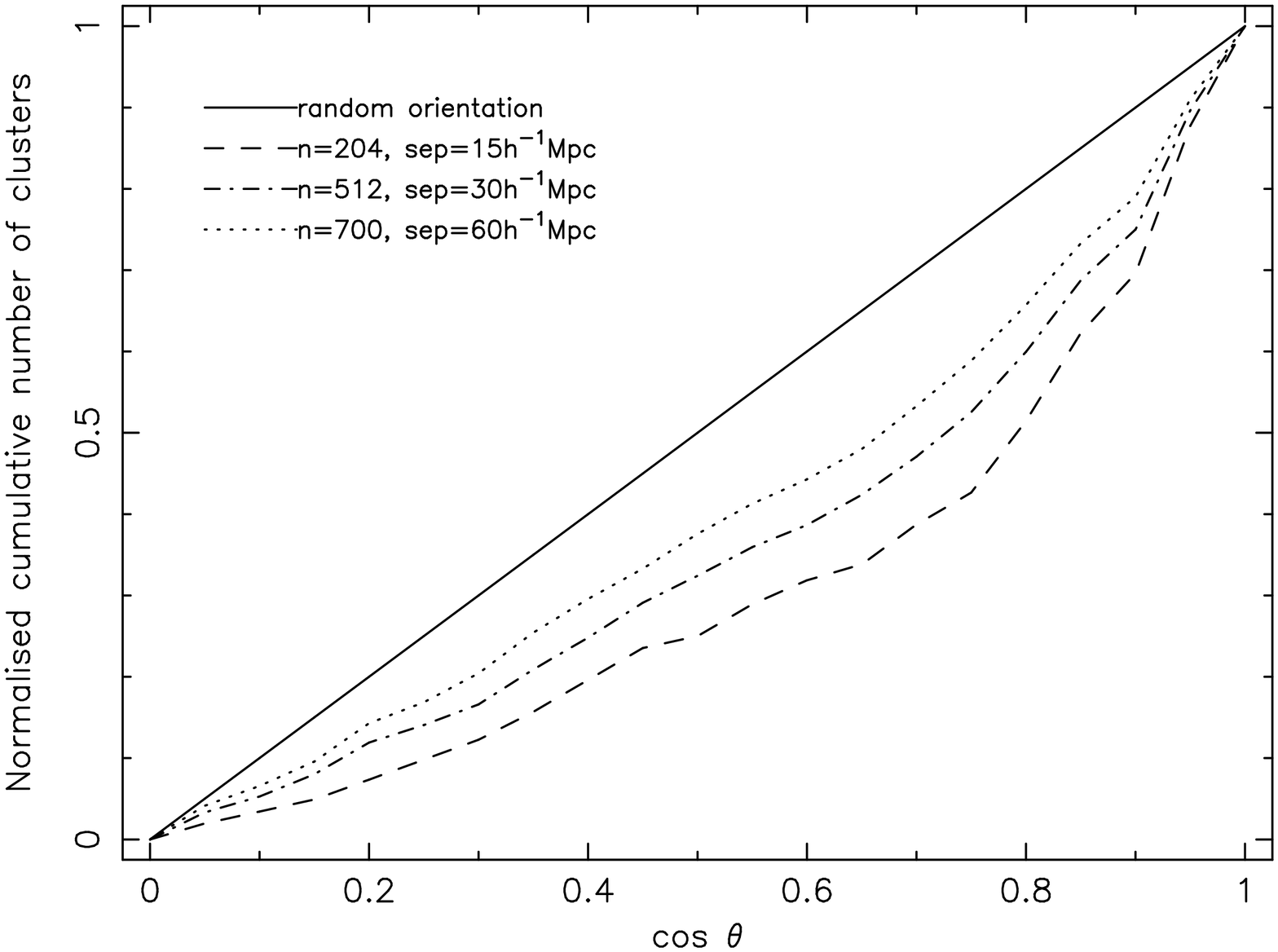,width=8.1cm,angle=270}}
\caption{Alignment of clusters with nearest neighbours for $\Lambda$CDM ($512^{3}$particles)}
\end{figure}

\begin{figure}
\psfig{figure=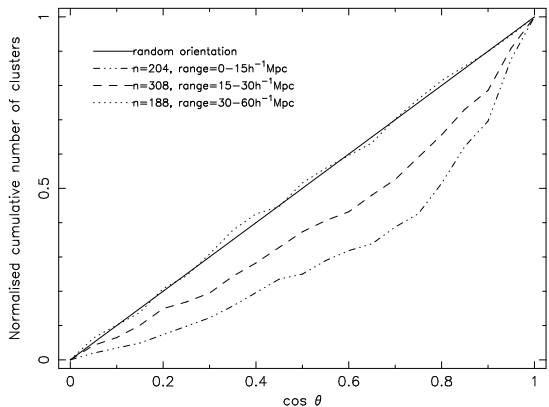,width=8.1cm,angle=270}
\caption{Alignment of clusters with nearest neighbours in separation ranges for $\Lambda$CDM ($512^{3}$ particles)}
\end{figure}
Table 3 and Figures 2 and 4 show that the alignments actually arise from neighbours closer than $60h^{-1}$Mpc, particularly in the $\tau$CDM model where almost all of the alignment arises from neighbours with separations of less than $15h^{-1}$Mpc. Unlike Splinter et al. (1997) and West, Vllumsen \& Dekel (1991), who found that the alignments extended to larger separations when clusters were limited to members of a supercluster, this situation was not found to change when cluster pairs within a supercluster were considered. The values of $d_{\mathrm{m}}$ remained very low at separations greater than $15h^{-1}$Mpc for the $\tau$CDM model. 
 
\begin{figure}
\psfig{figure=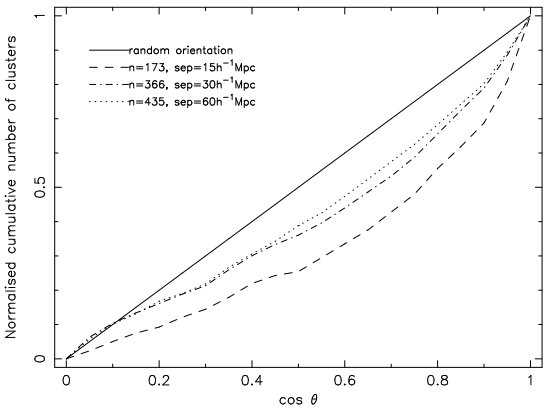,width=8.1cm,angle=270}
\caption{Alignment of clusters for nearest neighbours for $\tau$CDM ($512^{3}$ particles)}
\end{figure}

\begin{figure}
\psfig{figure=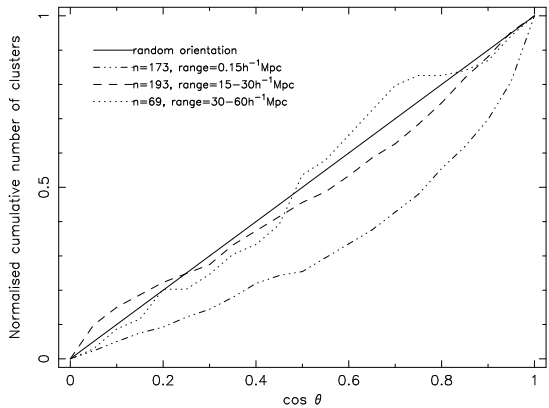,width=8.1cm,angle=270}
\caption{Alignment of clusters with nearest neighbours in separation ranges for $\tau$CDM ($512^{3}$ particles)}
\end{figure}

The significance of this difference in alignments found for the two cosmologies was investigated using a two-sided KS test. For the best discriminant between models, which was the separation range $15-30h^{-1}$Mpc, the KS probability of the alignments being the same in the two models was 0.14 with a maximum difference of 0.11. This is not a highly significant difference and is therefore
not likely to be useful for an observational test of cosmologies. However since smaller simulations were available for two other cosmological models it was decided to look for differences in alignments in this range ($15-30h^{-1}$Mpc) for all 4 models. 

\subsection{Alignments using $256^3$ particles for all four models}

The analysis carried out here with $512^{3}$ particles shows that there are some differences in the alignments for the $\Lambda$CDM and $\tau$CDM models particularly for separations between nearest neighbours in the range $15-30h^{-1}$Mpc. To check whether this could be used as a discriminator between other cosmological models the alignments in this range were also investigated for the $256^{3}$ particle simulations. The results of the alignments for different separation ranges for the $\Lambda$CDM, SCDM, OCDM and $\tau$CDM 
models are given in Table 4. The alignments in the range which was found to be the best discriminant for the larger simulations ($15-30h^{-1}$Mpc) are also shown in Figure 5 for all 4 models. 

\begin{table*}
\begin{center}
\Large
\begin{tabular}{|l|lcr|lcr|lcr|lcr|} \hline
Range & \multicolumn{3}{c}{$\Lambda$CDM} & \multicolumn{3}{c}{SCDM} & 
\multicolumn{3}{c}{OCDM} & \multicolumn{3}{c}{$\tau$CDM} \\ \cline{2-13}
 & $d_{\mathrm{m}}$ & p & n & $d_{\mathrm{m}}$ & p & n & $d_{\mathrm{m}}$ & p & n & $d_{\mathrm{m}}$ & p & n \\ \hline
 0-15 & $0.27$ & $2.9\times 10^{-2}$ & $29$ & $0.27$ & $1.2\times 10^{-4}$ & $69$ & $0.19$ & $0.21$ & $32$ & $0.32$ & $2.9\times 10^{-5}$ & $55$ \\
15-30 & $0.23$ & $0.14$ & $26$ & $0.08$ & $0.69$ & $91$ & $0.28$ & $1.3\times 10^{-2}$ & $33$ & $0.21$ & $1.1\times 10^{-2}$ & $61$ \\
30-60 & $0.07$ & $0.99$ & $33$ & $0.1$ & $0.93$ & $30$ & $0.16$ & $0.44$ & $28$ & $0.14$ & $0.82$ & $21$ \\
\hline
\end{tabular}
\end{center}
\caption{Alignments for $256^{3}$ nearest neighbours in separation ranges 0-15, 15-30 \& 30-60 $h^{-1}$Mpc for all 4 cosmological models }
\end{table*} 

\begin{figure}
\centerline{\psfig{figure=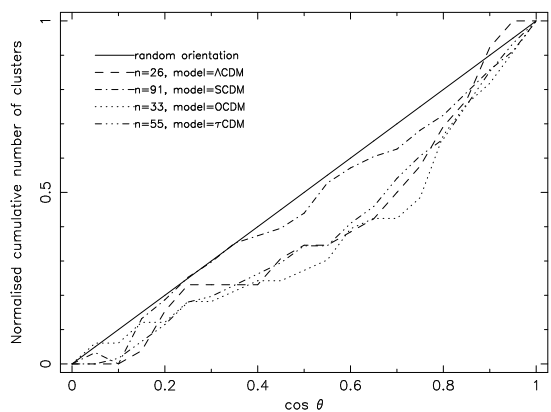,width=8.1cm,angle=270}}
\caption{Alignment of clusters in the separation range $15-30h^{-1}$Mpc
for nearest neighbours for 4 cosmological models}
\end{figure}

Differences in alignments between the cosmological models are not so clear cut for these smaller $256^{3}$ particle simulations, but this is not unconnected to the much smaller number of clusters in each sample. From Figure 5 it can be seen that the only model for which there are indications of a difference is the SCDM model, with no alignment arising at separations greater than $15h^{-1}$Mpc. This also is the only model which has a reasonably large number of clusters in the sample. The alignments cannot be used to discriminate between the other 3 samples since the numbers of clusters are so small that any variations are merely statistical fluctuations. It may be noted that the $\tau$CDM result is consistent with the result from the larger simulation given the poor number statistics of this $256^{3}$ simulation.

\section{DISCUSSION}

The results from the very large simulations confirm without doubt that alignments between clusters of galaxies do exist. This effect is stronger between nearest neighbours than for all neighbours, even if they are restricted to cluster pairs within a supercluster. A difference in the alignments was found for the two cosmological models investigated in that the alignments extended to larger separations (about $30h^{-1}$Mpc) for the $\Lambda$CDM model than for the $\tau$CDM model ($<15h^{-1}$Mpc). However the significance of this difference is not large enough to be used as an observational test. 

Splinter et al. (1997) found that there appears to be a very weak trend that as $\Omega$ is lowered more alignments are seen. They found a stronger trend that as the exponent of the primordial power spectrum (n) is made more negative there is increasing alignment between clusters. In the present simulations the power spectrum and shape parameter ($\Gamma$) were the same for both the $\Lambda$CDM and $\tau$CDM models. However in our simulations the normalization of the power spectrum ($\sigma_{8}$) was chosen to give the correct number of clusters at the present day for a particular cosmological model, giving  more power on large scales for low $\Omega$ models. It is thus not surprising that the alignments act over a longer range in the $\Lambda$CDM model.  

Although Binggeli (1982) found no relation between the orientation effect and cluster richness, we found some evidence that alignments increased with cluster richness. However, in agreement with Fuller et al. (1999), we found that alignments may persist down to poorer clusters (mass in region   $10^{13}-10^{14}\mathrm{M}_{\odot}$). 

In conclusion, it appears that cluster alignments are present for all CDM models  up to separations of $15h^{-1}$Mpc. The alignments extend to greater separations for the low $\Omega$ models at least, but the differences between models are not strong enough to be useful as a cosmological test. The alignments found may fit in with a general picture of cluster formation by hierarchical clustering in which material falls into the cluster along the large scale filamentary structure, possibly irrespective of cluster richness.         

\section{Acknowledgements}

The simulations analysed in this paper were carried out as part of the programme of the Virgo Supercomputing Consortium (http://star-www.dur.ac.uk/~frazerp/virgo/) using computers based at the Computing Centre of the Max-Planck Society in Garching and at the Edinburgh Parallel Computing Centre. 
LIO is in receipt of a Daphne Jackson Fellowship funded by the Royal Society.
PAT is a PPARC Lecturer Fellow.

\end{document}